\documentclass{epl}
\title{Localised and nonlocalised structures in nonlinear lattices with fermions}
\author{A. S. Carstea \inst{1}\footnote{email: acarst@theor1.theory.nipne.ro, 
carstas@yahoo.com}\and D. Grecu \inst{1} \and A. Visinescu \inst{1}}
\institute{
\inst{1} Department of Theoretical Physics, Institute of Physics and Nuclear
Engineering - Magurele, P.O. Box MG6 Bucharest, ROMANIA}
\begin{document}
\maketitle
\begin{abstract}
We discuss the quasiclassical approximation for the equations of motions 
of a nonlinear chain of phonons and electrons  
having phonon mediated hopping. 
Describing the phonons and electrons as even and odd grassmannian
functions and using the continuum limit we show that
the equations of motions lead to a Zakharov-like system for 
bosonic and fermionic fields.
Localised and nonlocalised solutions are discussed using the Hirota 
bilinear formalism. Nonlocalised
solutions turn out to appear naturally for any choice of wave parameters.  
The bosonic localised solution has a fermionic dressing 
while the fermionic one is an oscillatory localised field. They appear only
if some constraints on the dispersion are imposed. 
In this case the density
of fermions is a strongly localised travelling wave.   
Also it is shown that in the multiple scales approach the emergent equation
is {\it linear}. Only for the resonant case we get a nonlinear fermionic
Yajima-Oikawa system. Physical implications are discussed.
\end{abstract}

\section{Introduction}
The idea of treating classically systems with fermions goes back to seventies 
to the papers of Berezin \cite{1}, Casalbuoni \cite{2}. This is done using grassmannian
functions (elements of a Grassmann algebra). A simple representation can be done using,
for instance, combinations of Dirac matrices, namely 
$\alpha=\gamma_1+i\gamma_2, \beta=\gamma_3+i\gamma_4$. The algebra ge\-ne\-ra\-ted 
by $(1,\alpha,\beta, \alpha\beta)$ is a grassmann algebra with two generators.
The even sector (bosonic or commutative) is given by linear combinations of 
$(1,\alpha\beta)$, while the odd sector (fermionic or anticommutative) 
is given by linear combinations of $(\alpha, \beta)$.

Today, grassmannian description is used intensively
in mathematical physics. 
It is a very interesting idea because it could be 
implemented to nonlinear evolution equations containing 
bosons and fermions, for which the quantum description is difficult to do. Moreover, 
since the quantum mechanics deals mainly with observables and spectra, classical 
description offers an idea of the dynamics of solutions 
for the equations of motions. 
In this view there are some results 
related to the dynamics of solitons in integrable equations containing 
bosons and fermions but linked supersymmetrically \cite{3}. 
With the aid of a supersymmetric bilinear formalism these type of equations
have been solved. Interesting enough 
the existence of fermions makes the soliton interaction to
be nonelastic having a dressing in the fermionic part.    
  
Due to these new results it seems naturally to try to 
extend them to some physical  
models involving bosons and fermions and to see how the dynamics 
of the localised solutions is influenced by the fermions.

The main realm is of course the problem of localised structures in 
bio\-mo\-le\-cules (proteins, DNA, etc.). 
Since the seminal ideas of A. S. Davydov
\cite {5}, it has been continuing an intense quest 
for finding experimentally these 
nonlinear structures (solitons, breathers, etc.)\cite{peyrard}. In any case, 
insofar the results are controversial.

In this paper we consider the simplest  model namely 
a nonlinear chain 
and describe the 
phonons and electrons as functions having values in the even and the odd sector 
of an infinitely generated 
grassmann algebra (this can be done either directly or using coherent states). 
The choice of SSH hamiltonian is not only for  simplicity, but also it captures the
essential informations when one deals with a nonlinear lattice having a 
coupling between charge and structure (which explains charge conduction, 
polaron formation etc.) \cite{4},\cite{bishop}.

In the continuum limit the differential-difference 
equations of motions become a system which looks like the classical Zakharov
system \cite{6} with fermions. Interesting enough this system admits exact 
bilinearisation in the Hirota formalism and {\it two} type of solutions. 
One type is exponential and nonlocalised although it appears
to look like a kink type solution. But the presence of fermionic 
parameters at denominators breaks the localisation. The other type is 
localised even though is dressed with fermionic parameters as well. 
This type of solutions is responsible for a soliton-like 
propagating structure of the fermionic density in the chain.

\section{Equations of motion}
The hamiltonian we are working with is 
\begin{equation}
{\cal H}=\sum_n\frac{\dot u_n^2}{2}+W(u_{n+1}-u_{n})
+(c_{n+1}^{\dagger}c_n+c_n^{\dagger}c_{n+1})
\left[-t_0+\gamma(u_{n+1}-u_n)\right]
\label{1}
\end{equation}
where $u_n$ is the displacement from the equilibrium position of the particle 
$n$ and $c_n^{\dagger}$($c_n$) are the creation (annihilation) operators for
fermions on the site $n$. Also the potential $W(u_{n+1}-u_{n})$ can be harmonic 
as in the Su-Schriffer-Heeger (SSH) problem \cite{4} or anharmonic. 
We do not consider the spin indices because
we are not interested here in magnetic phenomena. The last term which contains 
the phonon mediated hopping is responsible for the interesting physics of the
system. 
The classical hamiltonian can be obtained by averaging the quantum one using  the 
product of the following coherent states (or directly by substitution)
$$|\chi_n>=e^{-\chi_n^{*}\chi_n/2}e^{\chi_n c_n^{\dagger}}|0>,$$
where $\chi_n(t)$ are grassmann-valued 
odd (anticommuting) complex functions. Also $u_n(t)$ are, from now on 
grassmann-valued even (commuting) functions (they are {\it not} 
real functions anymore). They belong to the even and odd sector of an 
infinitely generated grassmann algebra. Complex conjugation is defined 
such that $(z_1z_2)^{*}=z_2^{*}z_1^{*}$ in accordance with the conventions in
\cite{7}. As we are dealing with the classical case we assume further that
all fields and their derivatives commute or anticommute, depending in the 
usual way on the bosonic or fermionic nature of the fields.

First we consider the harmonic approximation, namely $W(u_{n+1}-u_{n})=K/2(u_{n+1}-u_{n})^2$. 
After averaging on the coherent states we find the following classical 
hamiltonian. 
\begin{equation}
H=\sum_n\frac{\dot u_n^2}{2}+\frac{K}{2}(u_{n+1}-u_{n})^2
+(\chi^{*}_{n+1}\chi_n+\chi^{*}_n\chi_{n+1})
\left[-t_0+\gamma(u_{n+1}-u_n)\right]
\label{2}
\end{equation}
The equations of motions are given by \cite{2}:
\begin{equation}
\ddot u_n=K(u_{n+1}+u_{n-1}-2u_n)+
\gamma(\chi^{*}_{n+1}\chi_n+\chi^{*}_{n}\chi_{n+1}-
\chi^{*}_{n}\chi_{n-1}-
\chi^{*}_{n-1}\chi_{n})\label{e1}
\end{equation}
\begin{equation}
i\dot\chi_n=-t_0(\chi_{n+1}+\chi_{n-1})
+\gamma\left[(u_n-u_{n-1})\chi_{n-1}+(u_{n+1}-u_n)\chi_{n+1}\right]\label{3}
\end{equation}
A tractable version of the above equations can be obtained using their 
continuum limit. 
Considering 
$x=\epsilon n$ we keep only terms up to order two in the expansions, i.e.
$u_{n\pm 1}=u(x,t)\pm u_x+u_{xx}/2$ and the same for $\chi_{n\pm 1}$. After
gauging the fermionic field $\chi\rightarrow \chi \exp(-2it_0t)$ and calling 
$K=c^2$ we find the following
system:
\begin{equation}
u_{tt}-c^2 u_{xx}=2\gamma(\chi^{*}\chi)_x\label{4a}
\end{equation}
\begin{equation}
i\chi_t+t_0 \chi_{xx}-2\gamma u_x\chi=0\label{4b}
\end{equation}
As expected we get the Zakharov system with bosonic and fermionic 
components.
The corresponding hamiltonian for these equations is given by:
\begin{equation}
H=\int dx\left(\frac{1}{2}u_t^2+\frac{c^2}{2}u_x^2+2\gamma\chi^{*}\chi u_x+t_0 \chi^{*}_x\chi_x\right)\label{h}
\end{equation}

\section{Localised and nonlocalised solutions}
In order to solve the equations of motions (\ref{4a}), (\ref{4b}) 
one can write all the quantities 
in terms of the generators of Grassmann algebra and then solve the equations 
layer by layer \cite{8}.  Even though the results can be very 
interesting, they depend 
on the number of generators. A different approach which is more appropriate for
computing the soliton-like solutions is the Hirota bilinear method \cite{9}. 
The idea behind it is to implement a nonlinear 
substitution for the field such that the nonlinear 
system is turned into a bilinear one having the form of the dispersion relation
when is written using the Hirota operators. On the other hand the nonlinear 
substitution somehow swallows the singularities of the solutions of nonlinear 
system and the bilinear one become a system which contains holomorphic
functions. In the case of integrable systems this fact is rigurously true. But
in the case of nonintegrable systems this works only for special solutions which, 
because of the holomorphic character, can be written in terms of exponentials.  
This strongly alleviates all the calculations. On the other hand the adaptation
of Hirota formalism to grassmannian functions is straightforward.

The nonlinear substitutions are:
$$\chi(x,t)=\frac{\sqrt{t_0}}{2\gamma}\frac{G}{F}, 
\quad u(x,t)=-\frac{t_0}{\gamma}\partial_x \log F$$ 
where $G(x,t)$ is an grassmann anticommuting complex function and $F$ is a
grassmann commuting real function.
Plugging into the nonlinear equations we find the following bilinear system:
\begin{equation}
({\bf D}_t^{2}-c^2{\bf D}_x^{2})F\bullet F=GG^{*}
\end{equation}
\begin{equation}
(i{\bf D}_t+t_0{\bf D}_x^{2})G\bullet F=0\label{5}
\end{equation}
where 
${\bf D}_x^{n}f\bullet g
=(\partial_\epsilon)^n f(x+\epsilon)g(x-\epsilon)|_{\epsilon=0}$
is the Hirota bilinear operator.
Now, usually the second bilinear equation gives the dispersion relation and 
the other one is only a constraint on the phases. 
The 1-soliton solution usually is given by the following ansatz,
\begin{equation}
G=\zeta e^{\eta}, \quad F=1+e^{\eta+\eta^{*}+\phi}\label{a}
\end{equation}
where $\zeta$ is a complex anticommuting 
grassmann parameter $(\zeta=\zeta_R+i\zeta_I)$,
$\eta=kx+\omega t$ where $k$ and $\omega$ are commuting complex grassmann
parameters and $\phi$ a constant phase. 
After plugging the ansatz into the equation we find the following 
relations for the parameters:
\begin{equation}
\left[(\omega+\omega^{*})^2-c^2(k+k^{*})^2\right]e^{\phi}
=\frac{1}{2}\zeta\zeta^{*}\label{7}
\end{equation}
\begin{equation}
i\omega+t_0 k^2=\alpha \zeta\zeta^{*}\label{8}
\end{equation}
The second equation and its complex conjugate 
arise as overall factors multiplicating terms containing $\zeta$ and 
$\zeta^{*}$. Accordingly these factors must be zero or proportional with the 
product $\zeta\zeta^{*}$ by a complex factor $\alpha$.

{\it Case $\alpha=0$}

\noindent As we pointed out, the eq. (\ref{8})
gives the dispersion relation of the soliton and the equation (\ref{7}) 
is necessary
to compute the phase $\phi$. Indeed, in our case,
$$e^{\phi}=\frac{\zeta\zeta^{*}}{2(\omega+\omega^{*})^2-2c^2(k+k^{*})^2}$$
i.e. is proportionally with the product of zetas. So, there are absolutely no restrictions
on $k$ and $\zeta$. 
For simplicity let's call the denominator
denominator $\Lambda^{-1}$.
Our 1-soliton solution will be (up to the factors)
$$\chi(x,t)=\frac{G}{F}=\frac{\zeta e^{\eta}}{1+\Lambda\zeta\zeta^{*} e^{\eta+\eta^{*}}},\quad u(x,t)=\frac{F_x}{F}=\frac{(k+k^{*})\Lambda \zeta\zeta^{*} e^{\eta+\eta{*}}}
{1+\Lambda\zeta\zeta^{*} e^{\eta+\eta^{*}}}$$
In the grassmann algebra we can compute exactly 
the inverse
of an commuting object, namely
$1/(a+b\zeta_1\zeta_2)=1/a-\zeta_1\zeta_2b/a^2$
and this is because of the fact that $\zeta_i^2=0$. 
Applying this rule to our soliton we shall find that our soliton is not 
at all a soliton but a pure 
{\bf nonlocalised} exponential solution.
\begin{equation}
\chi(x,t)=\zeta e^{\eta},\quad u(x,t)=(k+k^{*})\Lambda\zeta\zeta^{*} e^{\eta+\eta^{*}}
\end{equation}  
One can see that the bosonic solution is purely grassmannian 
(its square is zero), so we can speculate 
that it exists only at the quantum level. 
Accordingly, the usual method of finding localised 
solutions for Hirota bilinear
equations here fails due to the grassmannian character of 
the parameters.

There is a big drawback of these solutions. They have no
physical meaning since they grow
exponentially and the energy as well.

So we are forced to look for a method to find localised solutions. 
A posibility is to treat diferently  
equation (\ref{7}), namely to take it {\it as a dispersion relation} 
while the equation (\ref{8}) will give {\it constraints} on 
the solitonic parameters. 
For instance we can consider that $e^{\phi}$ is not at all 
proportional to the product of zetas but 
just a pure number (let's say 1/8)
and zeta's enter
in a the relation of $\omega$ and $k$ i.e
$$(\omega+\omega^{*})^2-c^2(k+k^{*})^2=e^{-\phi}\zeta\zeta^{*}/2$$
If we put $\omega=\omega_R+i\omega_I$, $k=k_R+ik_I$ and $e^{\phi}=1/8$
we find the following relation
$$\omega_R=\sqrt{c^2k_R^2+\zeta\zeta^{*}}=
ck_R+\frac{1}{2ck_R}\zeta\zeta^{*}=ck_R+\frac{i}{ck_R}\zeta_I\zeta_R$$
the last equations being consequences of the grassmannian character
of $\zeta$ and $\zeta^{*}$. Also the last equation is a {\it real} 
grassmannian quantity despite the $i$ factor because of the definition of 
complex conjugation $(z_1z_2)^{*}=z_2^*z_1^*$.
Plugging this relation into the equation (\ref{8}) one obtains
$$k_I=-\left(\frac{c}{2t_0}+\frac{\zeta\zeta^{*}}{4t_0k_R^2 c}\right),\quad \omega_I=t_0\left(k_R^2-\frac{c^2}{4t_0^2}-\frac{\zeta\zeta^{*}}{4t_0^2k_R^2}\right)$$
These are the constraints in the sense that $k_I$ and $\omega_I$ are no longer free.
So, instead of having four free parameters $k_R$, $k_I$, $\zeta_R$, $\zeta_I$ we have only 
three $k_R, \zeta_R$ and $\zeta_I$.
With all these relations one can write (after some algebra)
completely the 1-soliton solution,
\begin{equation}
u(x,t)=\frac{k_R}{2}sech^2 \eta_0\left(e^{2\eta_0}+\frac{1}{k_R c}e^{t\zeta\zeta^{*}}\right)
\end{equation}
\begin{equation}
\chi(x,t)=\frac{\zeta}{2}sech \eta_0 
\exp{\left[-i\frac{c}{2t_0}x-i(k_R^2t_0-\frac{c^2}{4t_0})t\right]}
\end{equation}
where $\eta_0=k_Rx+ck_R t$. 
One can see that the 
fermionic field $\chi$ is indeed a localised field
with an oscillating amplitude. 
The problem of localisability for the 
bosonic field is not so obvious because of the fermionic {\it dressing} in the last 
exponential. 
In any case expanding the exponential (taking into account that the square of 
$\zeta\zeta^{*}$ is zero) we see that the problematic part 
of $u(x,t)$ is $(t/k_Rc) sech^2 \eta_0$ which {\it has} finite limits as time goes
to infinity. So we have indeed a {\bf localised solution} which is expressed
using bosonic and fermionic parameters. The bosonic field contain a 
fermionic correction and the fermionic field is a soliton with oscillating
complex amplitude.
Here the physical interpretation is clear. This can be seen as a classical analog of the 
localised polaron wave function for the SSH problem \cite{bishop}. 
With our method we can in principle prove the existence of localised 
states. Using this solution one can compute the behaviour of the fermionic density, 
namely $c_n^{\dagger}c_n$. In the continuum limit this expression is given by
$$\chi^{*}(x,t)\chi(x,t)=\frac{\zeta^{*}\zeta}{4}sech^2 \eta_0$$
which shows the strongly localised character of the charge density 
travelling wave.
Plugging these solutions in the expression of the hamiltonian (\ref{h}) we find the energy:
$$H=\frac{2k_R t_0^2c^2}{3\gamma^2}+\frac{2t_0^2}{3\gamma^2}\left(2k_R+\frac{1}{k_R}\right)\zeta\zeta^{*}$$
One can see that the dressing of the bosonic solution and the fermionic field gives a purely grassmannian correction.

{\it Case $\alpha \neq 0$}

\noindent In this case, considering $\alpha=\mu+i\nu$  
the relations for $k_I$ and $\omega_I$ have the same structure. 
Only the factors of zetas are translated 
with some terms proportional with $\mu$ and $\nu$.

Of course, a big advantage of this model is that the underlying 
bosonic part is just a linear
Klein-Gordon equation, so we do not have solitonic phenomenology. 
We can do absolutely the same machinery for a nonlinear completely integrable 
bosonic equation with rich solitonic phenomenology.
For instance, we consider a fermionic extension of the Boussinesq equation 
\begin{equation}
u_{tt}-u_{xx}-6u_xu_{xx}-u_{xxxx}=(\chi\chi^{*})_x
\end{equation}
\begin{equation}
i\chi_t+\chi_{xx}-2u_x\chi=0\label{4}
\end{equation}
This system can be seen as a continuum version of the equations 
of motions for the hamiltonian (2.1) with the potential (up to coefficients)
$W(u_{n+1}-u_{n})=1/2(u_{n+1}-u_{n})^2+1/3(u_{n+1}-u_{n})^3$ 
for a weak phonon 
mediated hopping (we neglected the coefficients in the equations).
Then with the same nonlinear substitution we shall get:
\begin{equation}
({\bf D}_t^{2}-{\bf D}_x^{2}-{\bf D}_x^{4})F\bullet F=G^{*}G, 
\quad (i{\bf D}_t+{\bf D}_x^{2})G\bullet F=0\label{5}
\end{equation}
The ansatz for 1-soliton solution gives the following dispersion relations
\begin{equation}
\left[(\omega+\omega^{*})^2-(k+k^{*})^2-(k+k^{*})^4\right]e^{\phi}
=\frac{1}{2}\zeta^{*}\zeta,\quad i\omega+k^2=\alpha \zeta^{*}\zeta
\end{equation}
One can see that everything works in the same way except that $\Lambda^{-1}$ is translated
with $(k+k^{*})^4$.

\section{Multiple scales approach}

Even though the continuum limit used above is a {\it naive} one 
we can prove that using the
systematic multiple scales method \cite{10} one gets the unidirectional approximation
of the system (2.5), (2.6). 

In order to do this we assume that the competition between dispersion and nonlinearity  
occurs at large scales of space and time. Accordingly, we introduce the 
following stretched variables:
$$\xi=\epsilon(n-v_gt), \quad \tau=\epsilon^2 t$$
and consider that the fermionic complex field is just a slowly modulated 
fermionic wave,
$$\chi_n(t)=e^{i(kn-\omega t)}\sum_{j\geq1}\epsilon^{j}\Phi_j(\xi, \tau).$$
In the exponential, $\omega$ is exactly the dispersion relation of the 
{\it second} discrete equation from the system (3), namely 
$\omega=-2t_0 \cos{k}$. Also $v_g$ in the definition of the streched space
is the group velocity corresponding to this dispersion.
For the bosonic field we take,
$$u_n(t)=\sum_{j\geq1}\epsilon^{j} W_j(\xi, \tau)$$ 

Using the discrete equations (\ref{3}) 
we shall find at the order 
$\epsilon^3$
$$i\Phi_{1\tau}=-\omega_2 \Phi_{1\xi\xi}-2\gamma \cos k W_{1\xi}\Phi_1$$
where $\omega_2=d^2\omega/dk^2$.
Also from the first discrete equation (\ref{e1}) we shall find 
\begin{equation}
\epsilon^3(v_g^2-c^2)W_{1\xi\xi}+{\cal O}(\epsilon^4)=
2\gamma\epsilon^3\cos k (\Phi_1^{*}\Phi_1)_\xi+{\cal O}(\epsilon^4)\nonumber
\end{equation}
where $c^2$ is the phononic velocity. Assuming 
that we have the nonresonant case
namely $v_g\ne c$ we find
$$W_{1\xi}=\frac{2\gamma \cos k}{v_g^2-c^2}\Phi_1^{*}\Phi_1$$
Introducing this in the above equation and taking 
into account that the square of the $\Phi_1$ 
is zero we get a {\bf linear} equation
\begin{equation}
i\Phi_{1\tau}+\omega_2\Phi_{1\xi\xi}=0
\end{equation}
Accordingly, in the nonresonant case there is no 
localised solution. 

For the resonant case i.e. $v_g=c$ we have 
to change the scaling of the fermion field
$$\chi_n(t)=e^{i(kn-\omega t)}
\sum_{j\geq1}\epsilon^{j+1/2}\Phi_j(\xi, \tau)$$
which leaves unchanged the second equation but the first one will be
$$\epsilon^3 (v_{g}^2-c^2)W_{2\xi\xi}-2\epsilon^4 v_g W_{1\xi\tau}+{\cal O}(\epsilon^5)
=2\epsilon^4\gamma {\cos k} (\Phi_{1}^{*}\Phi_{1})_{\xi}+{\cal O}(\epsilon^5)$$  
Because $v_g=c$, finally we have the following system:
\begin{equation}
W_{1\tau}+\frac{\gamma}{v_g}\cos k(\Phi_1^{*}\Phi_1)=0
\end{equation}
\begin{equation}
i\Phi_{1\tau}+\omega_2\Phi_{1\xi\xi}+2\gamma {\cos k} W_{1\xi} \Phi_1=0
\end{equation}
which is indeed the uni-directional approximation of the Zakharov-type system,
the so called Yajima-Oikawa system \cite{11}. 
This equation has also a bilinear form 
and we can compute the solutions
in exactly the same way.

\section {Conclusions}
The dynamics of localised solutions in a nonlinear chain with fermions
described by SSH model.
is solved here using bilinear formalism adapted to a grassmannian
formulation of the equations of motions.  
The main problem here was that the usual way of finding solitonic
solutions gives purely exponential nonlocalised solutions. 
This fact happens even in the case of supersymmetric integrable equations 
\cite{12}
using the celebrated Darboux transform.  
Anyway including the grassmannian parameters into the dispersion relation
we have found really localised solutions but with some constraints on the parameters.    
So, one can speculate that 
in the nonlinear equations with fermions localised solutions exist only in special cases.

Also from the above multiple scale approach we can see that the existence
of localised solutions is somehow problematic at least for SSH model 
inasmuch the general 
nonresonant case in the standard scaling  goes to a linear equation for the 
fermionic field. 
Of course one can argue that a totally different scaling of space, time 
and fields might give a nonlinear equation for the fermionic field.
We do not believe this since purely fermions cannot selforganize into solitons
due to the exclusion principle. In order to appear localised solutions into 
a fermionic equation this one must be coupled with a bosonic one.

\end{document}